# Quantum Crossing Symmetry as Heart of Ghost Imaging

D. B. Ion[1], M. L. Ion[2] and L. Rusu[1]

[1]*Institute for Physics and Nuclear Engineering, Department of Fundamental Physics, POB MG 6, Bucharest, Romania* [2]*Bucharest University, Department for Nuclear Physics, POB MG 11, Bucharest, Romania*

**Abstract**: In this paper it is shown that the key to understanding the ghost imaging mystery are the crossing symmetric photon reactions in the nonlinear media. Then, an intuitive mechanism for the description of the ghost imaging in terms of the quantum mirror (QM) is presented. Moreover, we prove that the ghost imaging laws depend only on the energy-momentum conservation and not on the photons entanglement.
PACS: 11.10.-z; 42.50.-p; 03.65.Ud

In 1995, at Baltimore University, professor (Dr.) Yanhua Shih initiated ghost-imaging research [1], by using entangled photons. In that experiment, one photon passed through stenciled patterns in a mask to trigger a detector, and another photon was captured by a second detector. Surprisingly, an image of the pattern between the two detectors appeared which the physics community called ghost-imaging. Some definitions of ghost imaging (see refs. [1-4]) are as follows: (i) Ghost-imaging is a visual image of an object created by means of light that has never interacted with the object. (ii) Ghost imaging is an unusual effect by which an image is formed using light patterns that do not emanate from the target object. (iii) Ghost imaging, is a novel technique in which the object and the image system are on separate optical paths. (iv) "Ghost-imaging is similar to taking a flash-lit photo of an object using a normal camera. The image is formed by the photons that come out of the flash, bounce off the object, and then are focused through the lens onto photo-reactive film or a charge-coupled array. But, in this case, the image is not formed from light that hits the object and bounces back," Dr. Shih said. "The camera collects photons from the light sources that did not hit the object, but are paired through a quantum effect with others that did".

Here, we must underline that ten years ago, based on the crossing symmetry of the SPDC-photon reactions, the authors of the papers [5-7] presented a new interpretation of all the ghost phenomena. Then, they introduced the concept of SPDC-quantum mirror (QM) and on this basis they proved some important Quantum Mirror physical laws which can be of great help for a more deep understanding of the ghost imaging phenomena.

Klyshko explained the entangled two-photon imaging in a fictitious yet fascinating way [8]. In his view, the ghost image could be understood as a two-photon geometric optical effect by using the so-called "advanced wave interpretation." Basically, the light was considered that starts at one of the detectors, propagates backwards in time, reaches the SPDC crystal and then propagates forward in time towards the other detector. In this interpretation, the two-photon source of SPDC is treated as a spherical mirror [3].

This article is aimed at exploring and analyzing the quantum nature of ghost imaging. It is true that classical challenges have never stopped. Quantum? Classical? A hot debate (see Refs.[4]) is currently focused on ghost imaging. This article defends the quantum mechanical point of view based on quantum crossing symmetric photon interactions proposed ten years ago in the papers [5-7]. By this we show that the heart of the ghost imaging can be the "quantum crossing symmetric interactions" (quantum-CSI). Then, the Klyshko interpretation [8] of the ghost imaging is included in a more general and exact form in the quantum mirror mechanism presented here (see Fig. 1).

*Crossing symmetric SPDC-photon reactions*- In quantum field theory, crossing symmetry [9] is a symmetry that relates S-matrix elements. Interaction processes involving different kinds of particles can be obtained from each other by replacing incoming particles with outgoing antiparticles after taking the analytic continuation. The crossing symmetry applies to all known particles, including the photon which is its own antiparticle. For example, the annihilation of an electron with a positron into two photons is related to an elastic scattering of an electron with a photon by crossing symmetry. This relation allows to calculate the scattering amplitude of one process from the amplitude for the other process if negative values of energy of some particles are substituted.

$$\gamma + e^- \rightarrow e^- + \gamma \quad (1)$$
$$e^- + e^+ \rightarrow \gamma + \gamma \quad (2)$$

By examination, it can be seen that these two interactions are related by crossing symmetry. It could then be said that the observation of Compton scattering implies the existence of pair annihilation and predicts that it will produce a pair of photons.

The first experiments on ghost imaging were performed using a pair of entangled photons produced by spontaneous parametric down conversion (SPDC). In this process, a primary pump (p) photon is incident on a nonlinear crystal and produces the photons idler (i) and signal (s) by the reaction: $p \rightarrow s + i$. These photons are correlated in energy, momentum, polarizations and time of birth. Some of these features, such as energy and momentum conservations: $\omega_p = \omega_s + \omega_i$, $\vec{k}_p = \vec{k}_s + \vec{k}_i$ are exploited to match in diverse experiments. (e.g., Momentum conservation in the "degenerate" case when the idler and the signal photons acquire the same frequency leads to the production of a pair of simultaneous photons that are emitted at equal angles relative to the incident beam). Now, if the S-matrix crossing symmetry [9] of the electromagnetic interaction in the spontaneous parametric down conversion (SPDC) crystals is taken into account, then the existence of the direct SPDC process (see Fig.1)

$$\gamma_p \rightarrow \gamma_s + \gamma_i \quad (3)$$

will imply the existence of the following crossing symmetric processes

$$\gamma_p + \overline{\gamma}_s \rightarrow \gamma_i \quad (4)$$
$$\gamma_p + \overline{\gamma}_i \rightarrow \gamma_s \quad (5)$$

as real processes which can be described by the same transition amplitude. Here, by $\overline{s}$ and $\overline{i}$ we denoted the time reversed photons relative to the original photons s and i, respectively.

In fact, the SPDC-effects (4)-(5) can be identified as being directly connected with the $\chi^{(2)}$ – second-order nonlinear effects called in general three waves mixing. So, the process (3) is just the inverse of second-harmonic generation, while, the effects (4)-(5) corresponds to the emission of optical phase conjugated replicas in the presence of pump laser via three wave mixing. Here some remarks are necessary in connection with the entanglement. If the quantum entanglement is a quantum mechanical phenomenon in which the quantum states of two or more objects have to be described with reference to each other, even though the individual objects may be spatially separated, then the crossing symmetry of an interaction can be interpreted as a more general kind of entanglement.

*Ghost imaging via SPDC-crossing symmetric photon reaction-* The main purpose here is to obtain an answer to the basic question: Is ghost imaging mystery solved via the quantum mirror (QM) introduced in ref. [5-7]? So, we start with a the definition of the quantum mirror and some of its physical laws [5-7].

*SPDC-Quantum Mirror (QM)* (see Fig.1). A quantum mirrors is called SPDC-QM if is based on the quantum SPDC phenomena (3)-(5) in order to transform signal photons, characterized by $(\omega_s, \vec{k}_s, \vec{e}_s, \mu_s)$, into idler photons with $(\omega_p - \omega_s, \vec{k}_p - \vec{k}_s, \vec{e}_s^*, -\mu_s) \equiv (\omega_i, \vec{k}_i, \vec{e}_i, \mu_i)$.

Therefore, according to the schematic description from Fig.1, a SPDC-QM is composed from:

a high quality *laser pump* (p), *a transparent crystal* in which all the three photon reactions (3)-(5) can be produced, all satisfying the same energy-momentum conservation laws: $\omega_p = \omega_s + \omega_i$, $\vec{k}_p = \vec{k}_s + \vec{k}_i$.

In these conditions a new geometric optics can be developed on the basis of the concept of quantum mirrors (QM) as shown in refs. [5-7]. Hence, the laws of the plane quantum mirror (see Fig.4) and that of spherical quantum mirror (SQM) (see Fig. (3a,b)), observed in the ghost imaging experiments [2-3], are proved as natural consequences of the energy-momentum conservation laws. By the quantum mirroring mechanism (see Fig.1) the objects and their images can be considered as being on the same optical paths. Therefore, the key of ghost imaging mystery can be given by the electromagnetic crossing symmetric photon reactions (3)-(5). Indeed, in the case of ghost imaging observed in the papers [2,3] the ghost image is produced as follows (see Fig.1a): The image forms indirectly from the signal photons that come out of the flash, bounce off the aperture, and then are focused through the lens onto QM where they are transformed in idler photons **i_s** which are collected in the idler detector. So, in this case, the image is not formed directly from signal photons that hits the aperture and bounces back. The image is formed only by idler **i_s**-photons from QM-sources that did not hit the object, but are obtained via crossing symmetric photon reaction (4) and not via photon reaction (3). Therefore, the crossing symmetry is the heart of ghost imaging, ghost diffraction and ghost interference, phenomena.

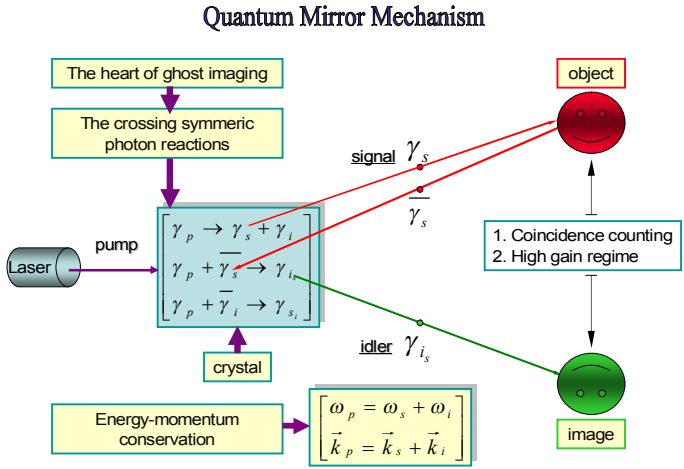

Fig.1: A schematic description of a SPDC-QM mechanism

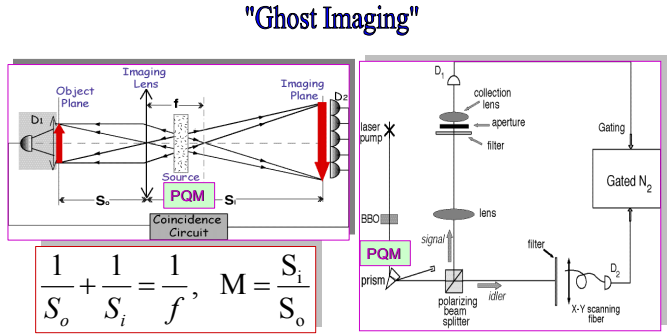

Fig.2: Law of the thin lens in the ghost imaging scheme where the PQM-source behaves just as a mirror.

Clearly, a SPDC crystal illuminated by a high quality laser beam can acts as real quantum mirrors since by the crossing processes (4) (or (5)) a signal photon (or idler photon) is transformed in an idler photon $i_s$ (or signal photon $s_i$), respectively. The quantum mirrors can be "plane"[7] (PQM) (see Fig.4) and "spherical quantum mirrors" (SQM) (see figs.3a,b) according with the character of incoming laser waves (plane waves or spherical waves). In order to avoid many complications, in this presentation we will work only in the thin crystal approximation. Now, it is important to note that using the QM-concepts [5-7] the results PQM and SQM from Fig.2-3 are obtained only as a consequence of the energy-momentum conservation (or phase matching conditions) without any kind of photon entanglement. In order to illustrate this we present in Fig. 3a proof of SQM-law using only energy-momentum conservation law.

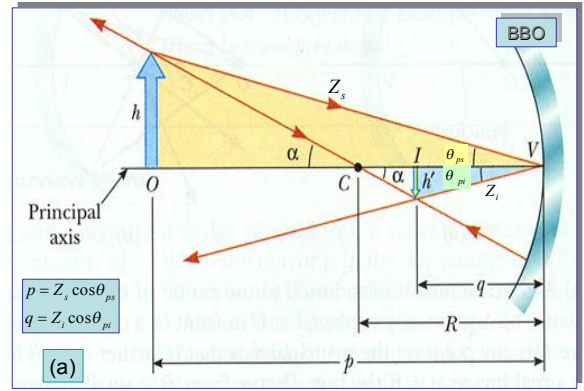

Fig.3. (a) A schematic description of the SQM; and (b) A short proof of the SQM-laws.

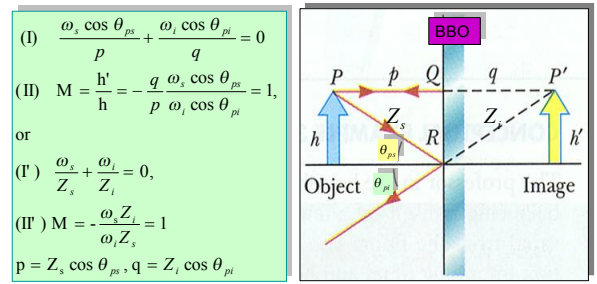

Fig.4: Laws of Plane Quantum Mirror

Also, it is important to remark that, the high quality of the SPDC-QM is given by the following peculiar characteristics (see Refs. [5-6]). *Coherence*: The SPDC-QM preserves high coherence between s-photons and i-photons; *Distortion undoing*: The SPDC-QM corrects all the aberrations which occur in signal or idler beam path; *Amplification*: A SPDC-QM amplifies the conjugated wave if some conditions are fulfilled; *Very high selectivity*: In the QM-mechanism the energy-momentum conservation law acts as a daemon which selects only the imaging photons $i_s$ which are produced by the crossing symmetric photon reaction (4).

*Conclusions*- The results obtained in this paper can be summarized as follows:

(i) The class of SPDC-phenomena (3) is enriched by introducing the crossing symmetric processes (4)-(5) as real phenomena described just by the same transition amplitude as that of the original SPDC-process (4) and satisfying the same energy-momentum conservation law (see Fig.1).

(ii) The QM-mechanism of the ghost-imaging is similar to taking a flash-lit photo of an object using a normal camera. The image forms indirectly from the signal photons that come out of the flash, bounce off the object, and then are focused through the lens onto QM where they are transformed in idler photons **$i_s$** which are collected on the photo-reactive film or a charge-coupled array. So, in this case, the image is not formed directly from signal photons that hits the object and bounces back. The image is formed only by idler **$i_s$**-photons from QM-sources that did not hit the object (see again Fig.1), but are obtained via crossing symmetric photon reaction (4) and not via photon reaction (3). In conclusion, the crossing symmetry is the heart of ghost imaging phenomena.

(iii) All the quantum mirror laws [5-7] (see Figs. 2-3 and [5-7]) are derived using only the energy-momentum conservation laws. The SQM-laws ($I_P$)-($II_P$) from Fig. 3b are verified experimentally with high accuracy by Pitman et al. [3]. Moreover, the recent results [10]-[12] definitely proved experimentally that the entanglement is not necessary in ghost imaging. Measurements in coincidence counting regime, as well as, in the high gain regime, can be used only for the background subtractions, just as in nuclear and elementary particle physics. Also, we remark that the results obtained in the paper [13], can be completely interpreted via the crossing symetric photon reactions (see Fig.1).

Finally, we think that the future goals must be to delve deeper into the quantum mirror physics of the ghost-imaging phenomenon, complete the theory of quantum mirroring, and to improve the technique toward practical QM-imaging applications and technologies.